\documentclass[11pt,aps,prd,showpacs,nofootinbib,superscriptaddress,preprint,preprintnumbers]{revtex4}

\pdfoutput=1

\usepackage{natbib}
\usepackage{color}
\usepackage{graphicx}
\usepackage{amsmath}
\usepackage[caption=false]{subfig}
\usepackage[colorlinks=true,linkcolor=blue,citecolor=blue,urlcolor=black]{hyperref}

\newcommand{\be}{\begin{equation}}
\newcommand{\ee}{\end{equation}}
\newcommand{\bea}{\begin{eqnarray}}
\newcommand{\eea}{\end{eqnarray}}

\usepackage{ulem,fancyvrb}

\usepackage{xcolor}
\usepackage{lipsum}

\begin{document}
\title{Degenerate Higgs bosons: hiding a second Higgs at 125 GeV}
\author{Ning Chen}
\affiliation{Department of Modern Physics, University of Science and Technology of China, Hefei, Anhui, 230026, China}
\author{Zuowei Liu}
\affiliation{School of Physics, Nanjing University, Nanjing, 210093, China}
\affiliation{Center for High Energy Physics, Tsinghua University, Beijing, 100084, China 
\\{
\smallskip 
\tt 
\href{mailto:chenning@ustc.edu.cn}{chenning@ustc.edu.cn}, 
\href{mailto:zuoweiliu@nju.edu.cn}{zuoweiliu@nju.edu.cn} 
\smallskip}},

\begin{abstract}

More than one Higgs boson may be present near the currently discovered Higgs mass, 
which can not be properly resolved due to the limitations in the intrinsic energy resolution at the Large Hadron Collider. 
We investigated the scenarios where two $CP$-even Higgs bosons are degenerate in mass. 
To correctly predict the Higgs signatures, quantum interference effects between the two Higgs bosons
must be taken into account, which, however, has been often neglected in the literature. 
We carried out a global analysis including the interference effects for a variety of Higgs searching channels at the Large Hadron Collider, which suggests that the existence of two degenerate Higgs bosons near 125 GeV is highly likely. 
Prospects of distinguishing the degenerate Higgs case from the single Higgs case are discussed.

\end{abstract}

\maketitle


\section{Introduction}

The discovery of the Higgs boson \cite{Higgs:1964ia, Higgs:1964pj, Englert:1964et} 
with mass close to 125 GeV 
at the Larger Hadron Collider (LHC) \cite{Aad:2012tfa, Chatrchyan:2012xdj} completes the particle spectrum 
of the standard model (SM) \cite{Weinberg:1967tq}. 
The Higgs boson in the SM has been extremely successful in explaining the LHC data in various  
Higgs boson search channels \cite{LHC:higgs}. 
Nonetheless, a richer structure beyond the simplest framework assumed in the SM 
may be present near 125 GeV, since one can not resolve the details of the spectrum near the Higgs boson mass with 
the current detection technology. 
The Higgs boson in the SM has a tiny decay width, about 4 MeV for mass near 125 GeV, 
which can not be directly measured via a scan on the lineshape of the Higgs boson resonance 
due to the limited energy resolution of the LHC. 
The Higgs boson decay modes with best energy resolution for mass reconstruction at the LHC are 
the ones with $\gamma\gamma$ and $ZZ\to \ell^+\ell^-\ell^{'+}\ell^{'-}$ as final states \cite{Aad:2015zhl}. 
The energy resolution in photon, electron, and muon is of the order of $\sim$GeV at the LHC
\cite{Aad:2014nim, Aad:2014rra, Chatrchyan:2012xi, Khachatryan:2015hwa, Khachatryan:2015iwa}, 
which is much larger than the total decay width of the SM Higgs boson.
Moreover, the details of the Higgs sector remain 
largely unknown; a number of well-motivated theoretical models predict a Higgs sector 
that contains more than one scalar field.

In this work, we investigate the possibility that two $CP$-even 
Higgs bosons have nearly degenerate masses near 125 GeV. 
Both Higgs bosons contribute to the observed signal events in 
all Higgs searching channels that have been collected by the 
ATLAS and CMS collaborations in their {$\sqrt{s}=7$ TeV, $\sqrt{s}=8$ TeV and $\sqrt{s}=13$ TeV} runs. 
Besides the individual contribution arising from the two Higgs bosons, 
the quantum interference effects of the 
two degenerate Higgs scalars must also be taken into consideration in order to correctly 
predict the signals at the LHC. 
To investigate the LHC signals, we take the two-Higgs-doublet model (2HDM) in which the degenerate Higgs 
scenarios can naturally arise as the prototype example, 
although our analysis can be extended to other models effortlessly.

The degenerate (or quasidegenerate) Higgs boson scenarios 
have been previously explored in the literature 
(see e.g.\ Refs.\ \cite{Gunion:2012gc, Gunion:2012he, Craig:2013hca}).
However, to our knowledge, the importance of the interference effects 
between the degenerate Higgs bosons 
have not been investigated before. 
In this work, we found that the quantum interference effects between  
the degenerate Higgs bosons can be rather substantial and can change the 
LHC signal predictions significantly. 
We further carried out a  
general global analysis in light of a variety of Higgs signal channels in 
 {the LHC runs with $\sqrt{s}=7$ TeV and $\sqrt{s}=8$ TeV,}
 \footnote{The current $\sqrt{s}=13$ TeV runs have only a few fb$^{-1}$ data analyzed, 
 which are less powerful in probing the Higgs boson(s) 
 at the $\sim$125 GeV scale than the 7 TeV and 8 TeV data, which have more than 20 fb$^{-1}$ 
 integrated luminosity combined, 
 since the ratios in cross sections between 13 TeV and 8 TeV LHC runs are generally in the range $(2-3)$.}
and found that the existence of two degenerate Higgs bosons near 125 GeV is highly likely.


\section{Two Higgs mediators in di-photon signal \label{sec:diphoton}}

\begin{figure}[htbp]
\vspace{0.2cm}
\centering
\includegraphics[width=0.5\textwidth]{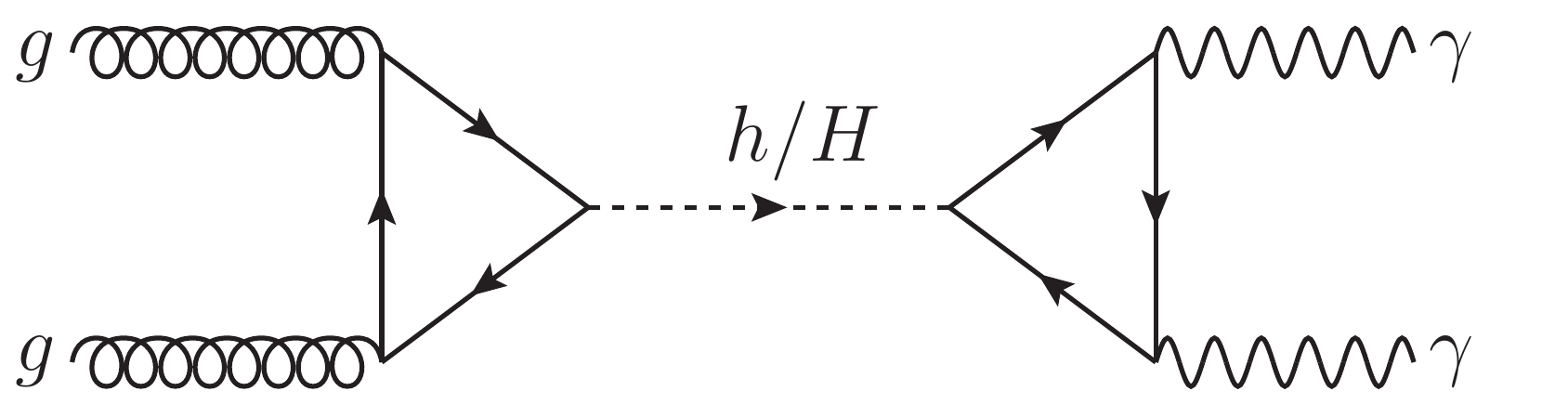}
\caption{Diphoton signal in the gluon fusion process. 
}
\label{fig:digram}
\end{figure}

To illustrate the effects of the additional Higgs boson, we consider 
the dominant 
channel of observing the Higgs boson at the LHC, the gluon fusion 
process (denoted as ggF) with two photons in the final states.
In the scenario where two $CP$-even Higgs bosons are degenerate in mass, 
both Higgs bosons enter as the s-channel mediators in the {ggF} process  
as shown {in} Fig.\ (\ref{fig:digram}).

We use $h$ to denote the lighter Higgs boson and $H$ the heavier one, 
with a small mass difference $\Delta M = M_H - M_h \geq 0$. 
To correctly calculate the cross section, we have to take into account both  the 
Breit-Wigner formula and the interference terms 
of the Higgs propagators. 
In the {ggF} process, $gg\to h/H \to \gamma\gamma$, 
the parton-level cross section is given by
\begin{equation}\label{eq:ggtoaa_parton_xsec}
\hat\sigma (gg\to h/H \to \gamma\gamma) \propto 
\left | \sum_{\phi=h,H} {G_{\phi gg} G_{\phi \gamma \gamma} 
\over \hat{s}-M_\phi^2 + i M_\phi \Gamma_\phi} \right|^2 \,,
\end{equation}
where $G_{\phi gg}$ and $G_{\phi \gamma \gamma}$  are the effective reduced couplings 
(dimensionless quantities with certain normalizations) 
of the Higgs boson $\phi$ to gluon and photon respectively, and   
$M_\phi\, (\Gamma_\phi)$ is the mass 
(total decay width) of the Higgs boson $\phi$. 
For the same process in the SM, only one Higgs boson contributes.

The total cross section at the LHC can be obtained via
$\sigma_\text{LHC}(s) = \int d\tau\,  \hat{\sigma}(\hat{s})\, d{\cal L}/d\tau (\tau)$,
where $\tau = \hat{s}/s$, $d{\cal L}/d\tau (\tau)$ is the parton luminosity. 
Because the Higgs decay widths are usually much smaller than the Higgs masses, 
i.e., $\Gamma_\phi$  $\ll M_\phi$, only the integration in the 
vicinity of the Higgs mass is significant.
The small variation in the 
parton luminosity near the Higgs mass can be safely neglected and 
the parton distribution functions
can be well approximated by a fixed value at the Higgs mass in this case.
Thus, the LHC cross section can be simplified as follows 
\bea\label{eq:ggtoaa_xsec}
\sigma_\text{LHC}(s) \propto  \int d\hat{s} \hat\sigma (\hat{s}) \propto 
\sum_{A,B=h,H} 
{ G_{A gg}  G_{A \gamma \gamma} 
G_{B gg}^* G_{B \gamma \gamma}^*  
\over  z_{A}- z^*_{B}}\,,
\eea
where $z_\phi \equiv M_\phi^2 - i M_\phi \Gamma_\phi$. 
To quantify the deviation from what expected in the SM, we normalize
the LHC cross section by its corresponding SM value as follows
\begin{equation}\label{eq:mu}
\mu
 = \sum_{A,B=h,H} {G_{Agg} G_{A \gamma\gamma} G^*_{B gg} G^*_{B \gamma\gamma} \over 
 |G_{h^0 gg} G_{h^0 \gamma\gamma} |^2} 
 {z_{h^0} - z_{h^0}^{*}  \over z_A-z_B^*}\,,
\end{equation}
where $h^0$ is the SM Higgs boson. 
The $\mu$ value here contains four terms: the $h$ contribution, 
the $H$ contribution, and the $h/H$ interference terms. 
Neglecting the interference terms, Eq.\ (\ref{eq:mu}) is reduced to a sum of two terms 
\be
\mu = {| \kappa_{h gg} \kappa_{h \gamma\gamma}|^2 \over \Gamma_h /\Gamma_{h^0} }
+{| \kappa_{H gg} \kappa_{H \gamma\gamma}|^2 \over \Gamma_H /\Gamma_{h^0} }\,,
\label{eq:kappa}
\ee
where we defined $\kappa_{\phi X} \equiv G_{\phi X} /G_{h^0 X}$ for any $X$ state, 
and we have neglected the small mass difference between the Higgs bosons. 
The $\kappa$ method given in Eq.\ (\ref{eq:kappa}) is often used in the literature to interpret experimental results 
for multiple Higgs boson scenarios. 
{However, since the $\kappa$ method ignores the important interference terms, 
it should be used with caution.}

\begin{figure}[htbp]
\vspace{0.2cm}
\centering
\includegraphics[width=0.5\textwidth]{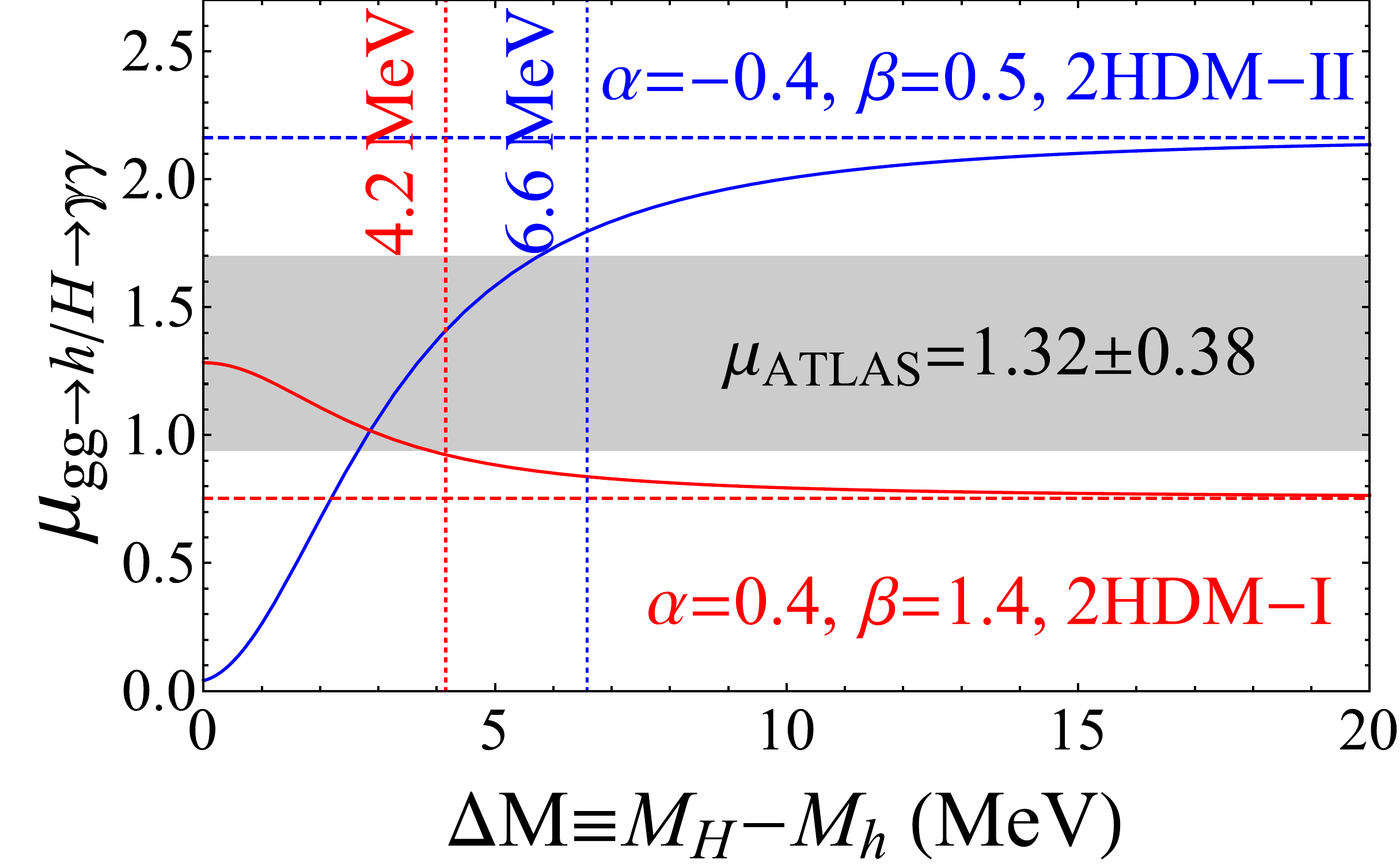}
\caption{(color online) Solid (dashed) curves are computed 
including (excluding) the interference terms. 
The gray band indicates the preferred $\mu$ range 
from ATLAS. 
Vertical dotted lines indicate the sum of the two Higgs decay widths, 
$\Gamma$. 
}
\label{fig:interference}
\end{figure}

To elucidate the effects due to interference and also the limitation of the $\kappa$ method, 
signal strengths computed using both Eq.\ (\ref{eq:mu})  and Eq.\ (\ref{eq:kappa})
for two model points are shown in Fig.\ (\ref{fig:interference}). 
For both models, the naive $\kappa$ method predicts a $\mu$ value disfavored by the 
ATLAS data, as shown by the two horizontal dashed lines in Fig.\ (\ref{fig:interference}). 
However, by including the interference effects, both models can actually give rise to 
a $\mu$ value within the ATLAS band for $\Delta M$ in the several MeV range.  
Fig.\ (\ref{fig:interference}) shows that the interference effects are important for 
$\Delta M\lesssim \Gamma\equiv \Gamma_h+\Gamma_H$ 
(we will use $\Gamma$ to denote the sum of the two Higgs decay widths throughout the paper) 
and can be either constructive as in the type-I 2HDM (hereafter 2HDM-I) model with $\alpha=0.4$ and $\beta=1.4$
or destructive as in the type-II 2HDM (hereafter 2HDM-II) model with $\alpha=-0.4$ and $\beta=0.5$.

\begin{figure*}[bthp]
\vspace{0.2cm}
\centering
\includegraphics[width=0.45\textwidth]{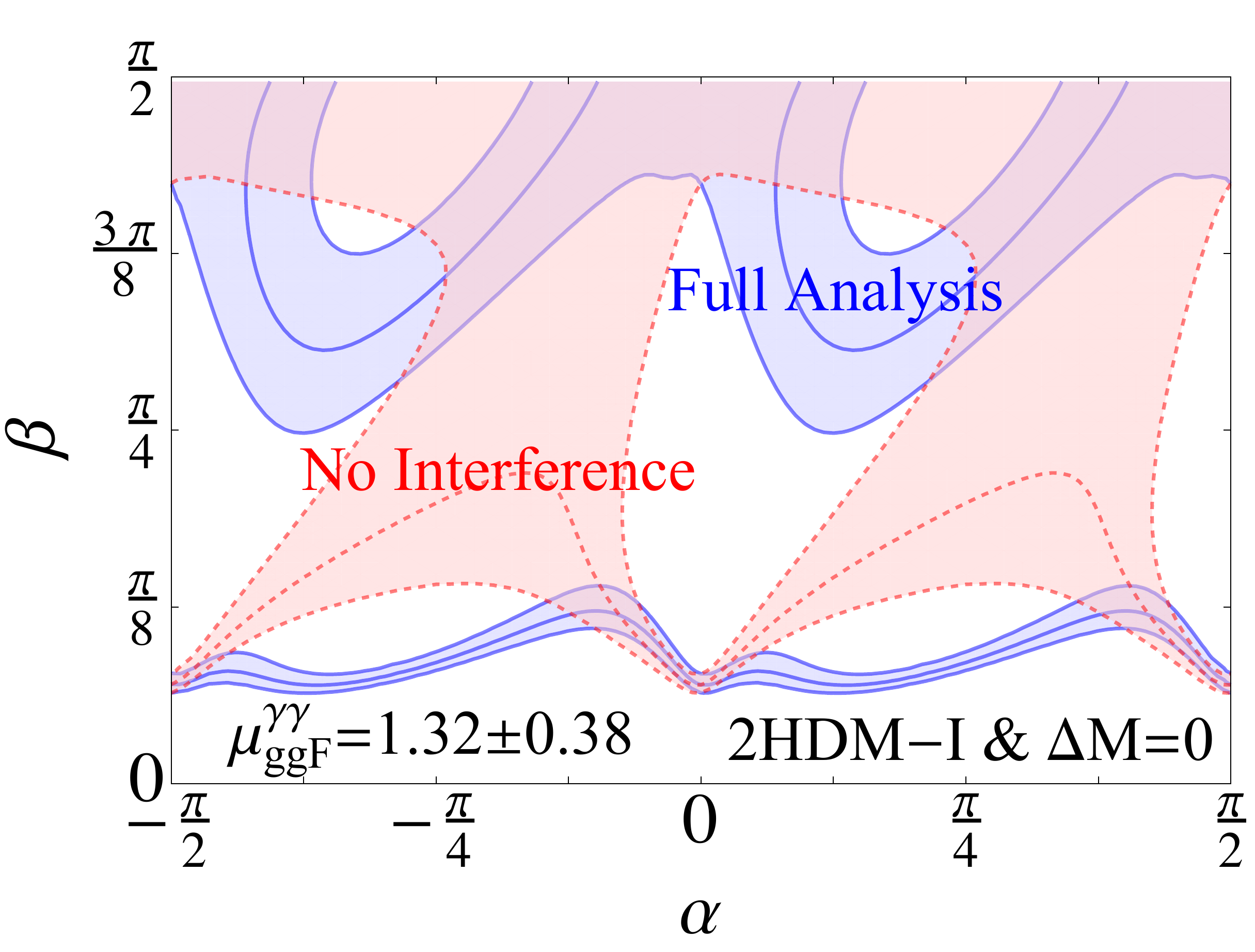}
\hspace{1cm}
\includegraphics[width=0.45\textwidth]{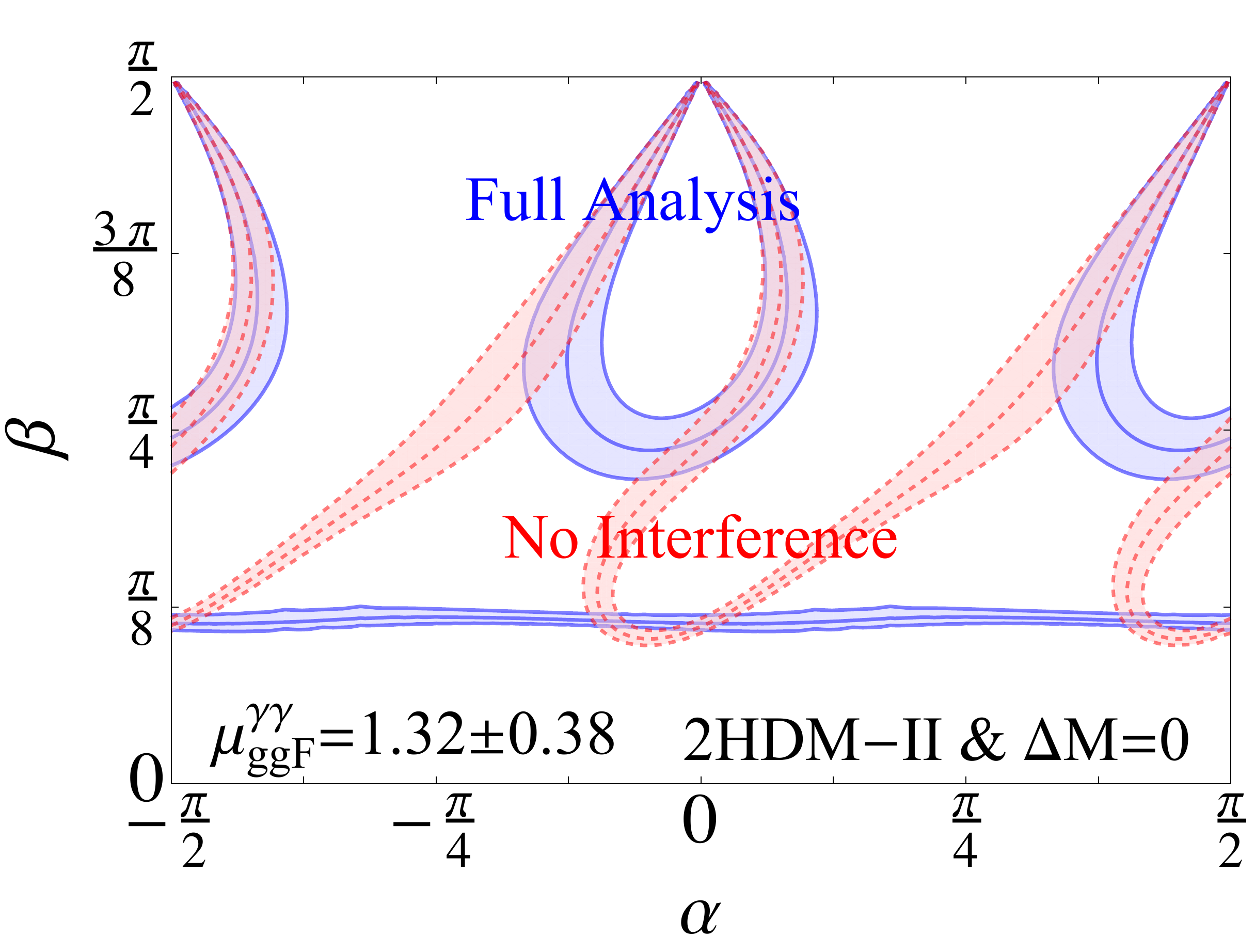}
\caption{(color online) 
Contours in the 2HDM-I (left) and the 2HDM-II (right) 
for the ATLAS measurement $\mu_\text{ggF}^{\gamma\gamma}=1.32\pm0.38$, 
in the $\Delta M = 0$ case. The blue region enclosed by 
blue solid boundaries are the full calculation including the interference terms between 
the two Higgs bosons; the red region enclosed by red dashed boundaries are computed 
without the interference terms. 
}
\label{fig:intercom}
\end{figure*}

In Fig.\ (\ref{fig:intercom}), the contours computed via  Eq.\ (\ref{eq:mu}) 
and Eq.\ (\ref{eq:kappa}) are drawn to fit the signal strength for diphotons via the gluon fusion processes measured by the ATLAS detector.
The two methods only overlap for a small fraction of the parameter space, as shown in Fig.\ (\ref{fig:intercom}).
Clearly, the $\kappa$ method can result in the parameter space where the 
signal strength has already been excluded by the ATLAS data, for example, 
the region near {$\alpha= \pm \frac{\pi}{4} $} and $\beta={\pi \over4}$ in the 2HDM-I as 
shown in {the left panel of} Fig.\ (\ref{fig:intercom}). 
Thus, analysis with the $\kappa$ method is inadequate in dealing with 
two degenerate Higgs bosons, since it can lead to erroneous conclusions.


\begin{table}[htbp]
\vspace{0.2cm}
\begin{center}
\begin{tabular}{c|c|c|c|c|c}
\hline
D & P &\multicolumn{4}{c}{$\mu\pm \sigma $} \\\hline
 &  & ATLAS & Ref & CMS & Ref   \\\hline\hline
$\gamma\gamma$ & ggF     & $1.32\pm 0.38$                    &\cite{Aad:2014eha}            & $1.12^{+0.37}_{-0.32}$         &\cite{Khachatryan:2014ira}           \\\hline
$\gamma\gamma$ & VBF    & $0.8\pm 0.7$                        & \cite{Aad:2014eha}           & $1.58^{+0.77}_{-0.68}$         & \cite{Khachatryan:2014ira}             \\\hline
$\gamma\gamma$ & WH     & $1.0\pm 1.6$                        & \cite{Aad:2014eha}           &-                                              &-                                                           \\\hline
$\gamma\gamma$ & ZH      & $0.1^{+3.7}_{-0.1}$              & \cite{Aad:2014eha}           &-                                              &-                                                      \\\hline
$\gamma\gamma$ & VH      &-                                            &-                                          & -$0.16^{+1.16}_{-0.79}$         & \cite{Khachatryan:2014ira}            \\\hline
$\gamma\gamma$ & ttH      & $1.6^{+2.7}_{-1.8}$              & \cite{Aad:2014eha}            & $2.69^{+2.51}_{-1.81}$          & \cite{Khachatryan:2014ira}           \\\hline\hline
$ZZ$ & ggF,ttH,bbH             & $1.7^{+0.5}_{-0.4}$              & \cite{Aad:2014eva}            &-                                             & -                                                              \\\hline
$ZZ$ & ggF,ttH                     & -                                           &-                                          & $0.80^{+0.46}_{-0.36}$            &\cite{Chatrchyan:2013mxa}             \\\hline
$ZZ$ & VBF,VH                    & $0.3^{+1.6}_{-0.9}$             &\cite{Aad:2014eva}             & $1.7^{+2.2}_{-2.1}$                     &\cite{Chatrchyan:2013mxa}              \\\hline\hline
$WW$ & ggF                        & $1.02^{+0.29}_{-0.26}$       &\cite{ATLAS:2014aga}         & $0.74^{+0.22}_{-0.20}$          &\cite{Chatrchyan:2013iaa}            \\\hline
$WW$ & VBF                       & $1.27^{+0.53}_{-0.45}$       & \cite{ATLAS:2014aga}        & $0.60^{+0.57}_{-0.46}$         &\cite{Chatrchyan:2013iaa}           \\\hline
$WW$ & VH                         &-                                            &-                                           & $0.39^{+1.97}_{-1.87}$          & \cite{Chatrchyan:2013iaa}           \\\hline\hline
$bb$ & ttH                            & $1.5\pm 1.1$                        &\cite{Aad:2015gra}              & $1.2^{+1.6}_{-1.5}$                    & \cite{Khachatryan:2015ila}           \\\hline
$bb$ & VH                           & $0.51^{+0.40}_{-0.37}$         & \cite{Aad:2014xzb}            & $1.0\pm 0.5$                               &\cite{Chatrchyan:2013zna}          \\\hline\hline
$\tau\tau$ & ggF                  & $2.0^{+1.5}_{-1.2}$               & \cite{Aad:2015vsa}            & $1.07\pm 0.46$                    & \cite{Chatrchyan:2014nva}           \\\hline
$\tau\tau$ & VBF,VH           & $1.24^{+0.59}_{-0.54}$          &\cite{Aad:2015vsa}             &-                                             &-                                                      \\\hline
$\tau\tau$ & VBF                 &-                                              &-                                          & $0.94 \pm 0.41$                   &\cite{Chatrchyan:2014nva}            \\\hline
$\tau\tau$ & VH                  &-                                               &-                                          & -$0.33 \pm 1.02$                    &\cite{Chatrchyan:2014nva}           \\\hline\hline
\end{tabular}
\caption{Signal strengths of Higgs searches measured by the ATLAS and  CMS collaborations, 
 for various decay and production channels for the $\sqrt{s}=$7 TeV and $\sqrt{s}=$8 TeV runs.
}
\label{tab:lhc}
\end{center}
\end{table}


\section{Global analysis \label{sec:global}}

To analyze the limits from LHC data, we carry out a global $\chi^2$ analysis 
for a variety of Higgs searching channels.
We categorize the experimental data in terms of the Higgs production channels (hereafter P) and 
Higgs decay  final states (hereafter D). 
There are six major production channels: 
$P$=(ggF, VBF, WH, ZH, ttH, bbH), 
where VBF stands for the vector boson fusion processes, including both WBF and ZBF, 
and the last four processes are the ones in which Higgs is produced in association with other particle(s) as indicated. There are five major decay final states used to detect the Higgs boson, 
which are $D$=($\gamma\gamma, ZZ, WW, bb, \tau\tau$). 
The $ZZ$ and $WW$ final states are usually tagged with $4\ell$ and $2\ell 2\nu$, respectively. 
A list of the major experimental results combining the 7 TeV data and the 8 TeV data in various Higgs searching 
channels from both ATLAS and CMS collaborations,  are summarized in Table (\ref{tab:lhc}). 
The experimental results are expressed in term of the signal strength, 
which is the ratio between the observed LHC signal events and its SM expectation.

The signal strength in the 2HDM for given production and decay channel 
(denoted by $\mu_\text{PD}$) can be computed via 
\begin{equation}\label{eq:muPD}
\mu_\text{PD}
 = \sum_{A,B=h,H} {G_{AL} G_{AD} G^*_{BL} G^*_{BD} \over 
 |G_{h^0 L} G_{h^0D} |^2} 
{z_{h^0} - z_{h^0}^{*}  \over z_A-z_B^*}\,,
\end{equation}
where $G_{AL}$ is the reduced coupling constant between the 
mediator $A$ and the state $L$, and so on.   
For the production channels $P$=(ggF, ttH, bbH), 
the corresponding $L$ states are $L=(gg, tt, bb)$; 
for $P$=(VBF, WH, ZH),  we have $L=VV$ in the 2HDM~\cite{custodial}.
{For completeness, we prove the validation of using Eq.~\eqref{eq:muPD} for the $WH$ and $ZH$ processes in the Appendix.}

The reduced Higgs couplings to up-type quarks and to $W$ and $Z$ bosons are the same for 
both 2HDM-I and 2HDM-II: $G_{huu} = \cos\alpha/\sin\beta$ and 
$G_{Huu} = \sin\alpha/\sin\beta$; 
$G_{hVV}=\sin(\beta-\alpha)$ and 
$G_{HVV}=\cos(\beta-\alpha)$. 
For down-type quarks, 
$G_{hdd} = \cos\alpha/\sin\beta$ and 
$G_{Hdd} = \sin\alpha/\sin\beta$ for 2HDM-I; 
$G_{hdd} = -\sin\alpha/\cos\beta$ and 
$G_{Hdd} = \cos\alpha/\cos\beta$ for 2HDM-II. 
The charged leptons have the 
same reduced coupling constant as the down-type quarks. 
For a review on the 2HDM, see e.g.\ Refs.\ \cite{Djouadi:2005gi, Djouadi:2005gj, Branco:2011iw}.  
In the SM, we have $G_{h^0X}=1$ for all the above couplings.

For Higgs couplings to gluon and photon, loop contributions via fermions and vector bosons
are calculated. The reduced couplings to gluons for both $h$ and $H$ in the
2HDM and for $h^0$  in the SM are given by 
$G_{\phi gg} = G_{\phi uu} A_{1/2}(x_{\phi t})+
G_{\phi dd} A_{1/2}(x_{\phi b})$ 
where $A_{1/2}(x)$ is the fermion loop function, 
$x_{\phi p} \equiv M_\phi^2/(4M_p^2)$ with $M_p$ being the mass of the 
circulating particle. The reduced couplings to photons are given by 
$G_{\phi \gamma\gamma }=N_c Q_u^2 G_{\phi uu} A_{1/2}(x_{\phi t})+
N_c Q_d^2 G_{\phi dd} A_{1/2}(x_{\phi b})
+ G_{\phi VV} A_{1}(x_{\phi W})$ 
where $N_c=3$, $Q_u=2/3$, $Q_d=-1/3$, and 
$A_{1}(x)$ is the vector boson loop function. 
The loop functions $A_{1/2}(x)$ and $A_{1}(x)$ can be found in e.g.\ Refs.\ 
\cite{Djouadi:2005gi, Djouadi:2005gj}. 
Here we include the bottom quark loops, because they can become 
significant in the 2HDM-II for large $\tan\beta$.

For the SM Higgs boson decay width, we use $\Gamma_{h^0}=4.07$ MeV  as the 
total width for $M_{h^0}=125$ GeV \cite{Denner:2011mq}. 
The partial decay widths are obtained by 
multiplying the total width with the following branching ratios: 
BR$(bb, \tau\tau, \mu\mu, cc, ss) = 
(5.77\times 10^{-1}, 6.32 \times 10^{-2}, 2.19 \times 10^{-4}, 2.91 \times 10^{-2}, 2.46 \times 10^{-4})$
for fermions, and 
BR$(gg, \gamma\gamma, Z\gamma, WW, ZZ) = 
(8.57\times 10^{-2}, 2.28 \times 10^{-3}, 1.54 \times 10^{-3}, 2.15 \times 10^{-1}, 2.64 \times 10^{-2})$ 
for bosons \cite{Denner:2011mq}. 
The partial decay widths of $h$ and $H$ in the 2HDM 
are computed via $\Gamma(\phi\to X) = \Gamma(h^0 \to X) |G_{\phi X}/G_{h^0X}|^2$ 
for any $X$ final state.

\begin{figure*}[thbp]
\vspace{0.2cm}
\centering
\includegraphics[width=0.45\textwidth]{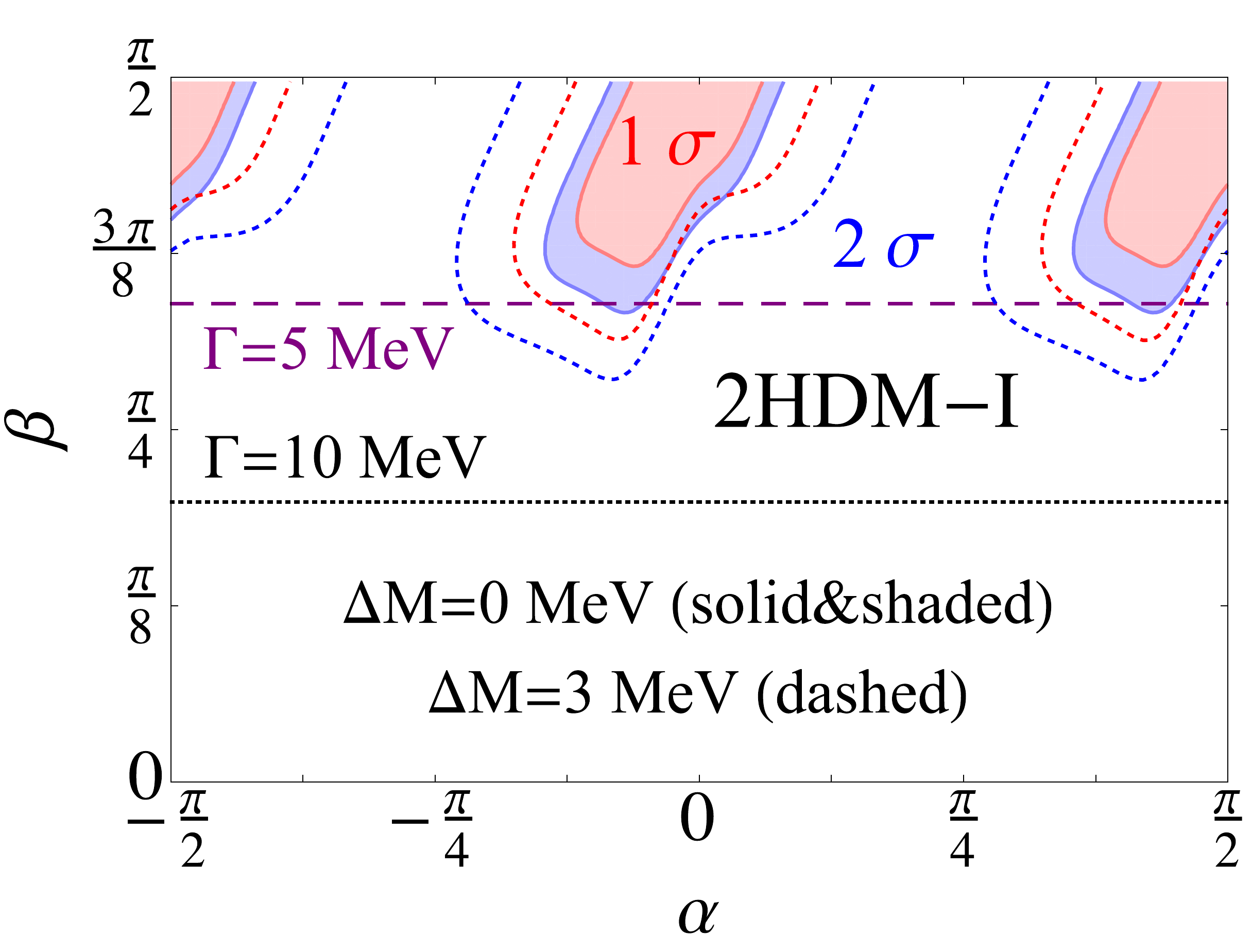}
\hspace{1cm}
\includegraphics[width=0.45\textwidth]{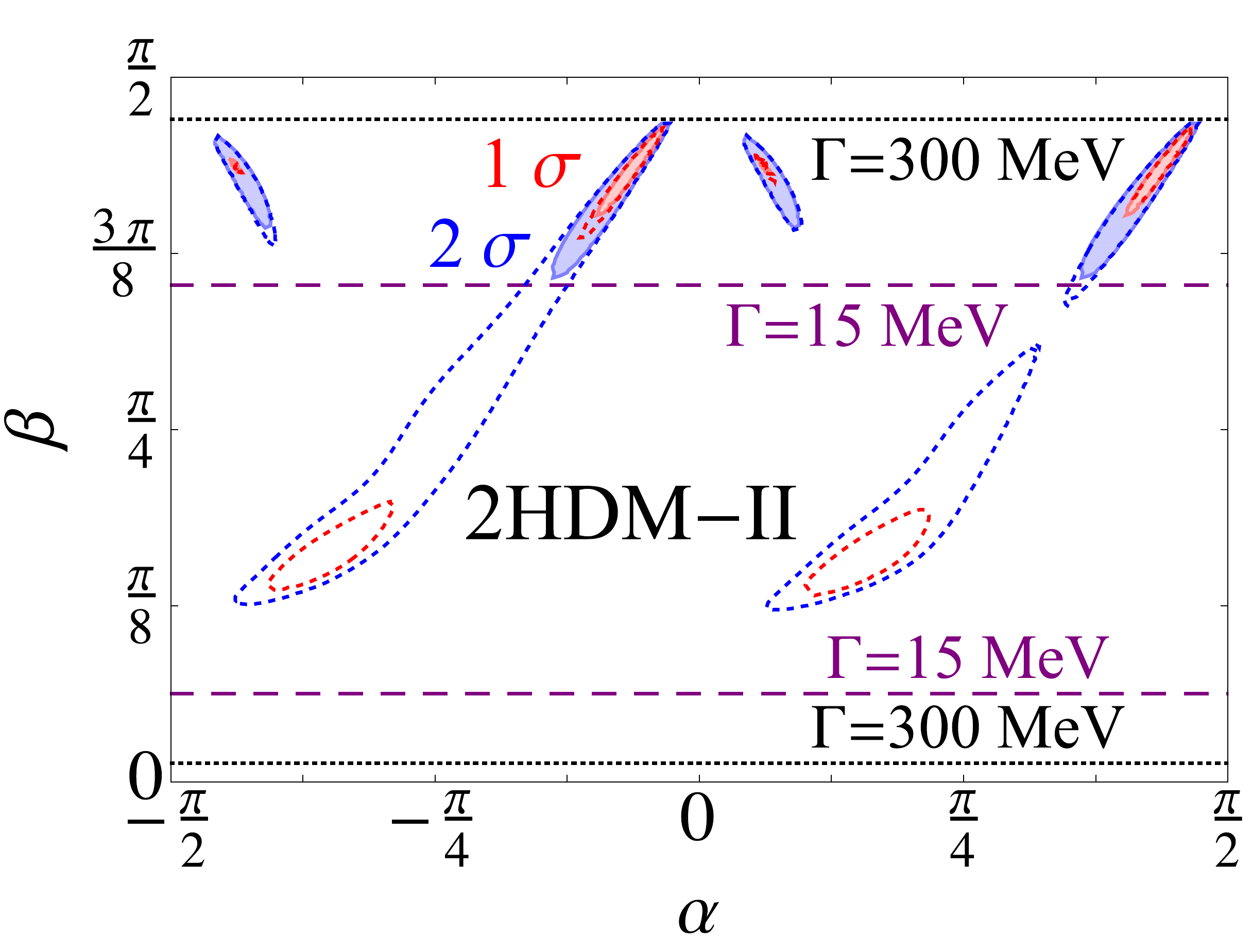}
\caption{(color online) 
Contours of $\chi^2_\text{LHC}$ in the 2HDM-I (left) and 
the 2HDM-II (right)  for $\Delta M =0$ (solid \& shaded) and $\Delta M =3$ MeV (dashed) cases. 
The horizontal lines indicate the sum of the decay widths $\Gamma$.
}
\label{fig:chi2}
\end{figure*}

\begin{figure}[htbp]
\vspace{0.2cm}
\centering
\includegraphics[width=0.5\textwidth]{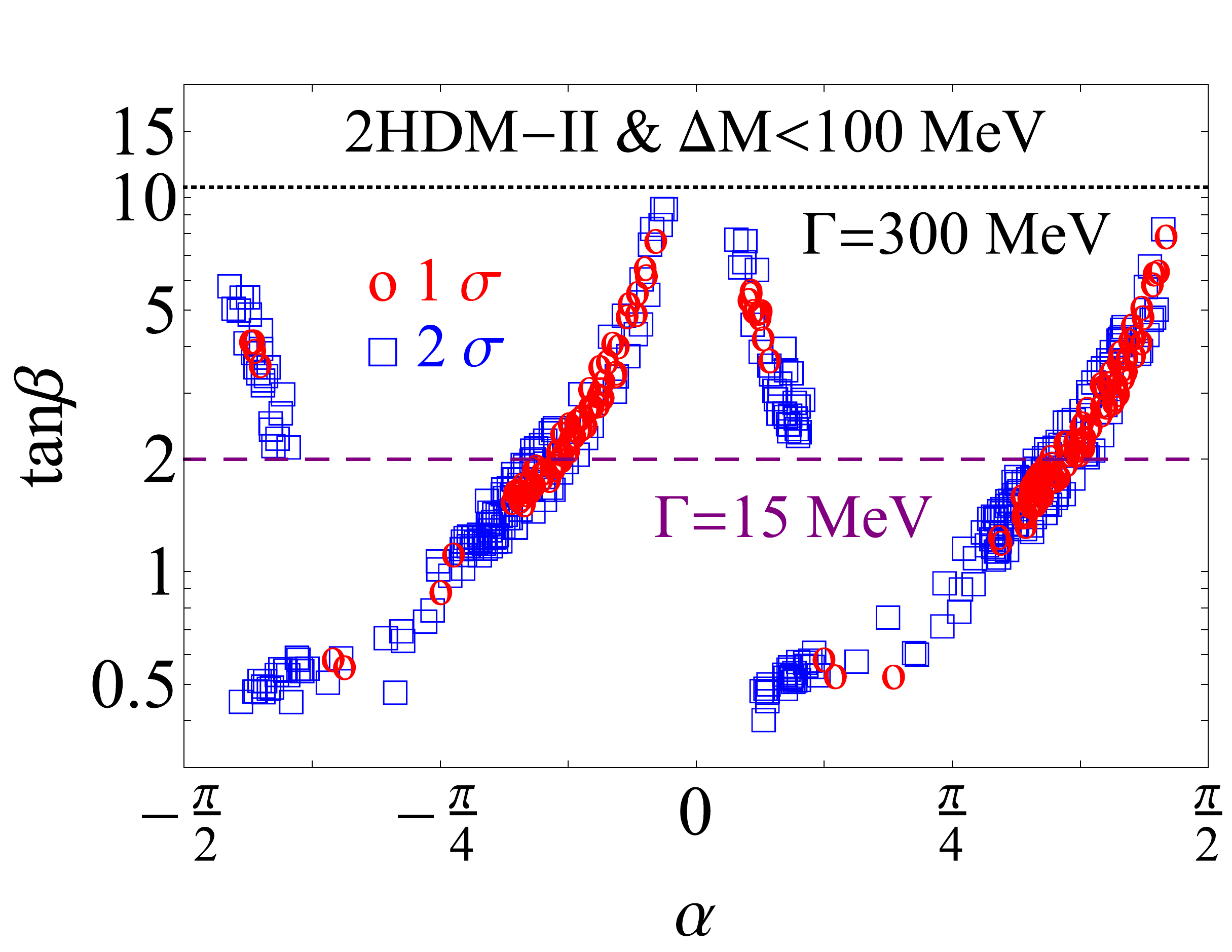}
\caption{(color online) 
Models within 1 $\sigma$ and 2 $\sigma$ 
corridors in $\chi^2$ near the best-fit model for the $\Delta M < $ 100 MeV case. 
}
\label{fig:3d}
\end{figure}

For most ATLAS and CMS analyses given in Table (\ref{tab:lhc}), the $\mu$ value for one single production channel is provided.
There are also signal strengths for multiple production channels.
For such analyses involving several production channels, we compute the signal strength as follows 
\begin{equation}
\mu_\text{multiple} = \sum_P \mu_\text{PD} N^\text{SM}_\text{PD} \Big/ \sum_P N^\text{SM}_\text{PD}, 
\label{eq:muth}
\end{equation} 
where $N^\text{SM}_\text{PD}={\cal L}_7\sigma^\text{SM-7}_\text{PD} (A\epsilon)^7_\text{PD}
+{\cal L}_8\sigma^\text{SM-8}_\text{PD}(A\epsilon)^8_\text{PD}$ 
is the expected number of events for the SM case at the LHC. 
Here $\sigma^\text{SM-7}_\text{PD}$ and ${\cal L}_7$ are the SM production cross section 
and the integrated luminosity at the $\sqrt{s}=7$ TeV LHC, and so on. 
In analyses combining the VBF and VH processes, we have $\mu_\text{multiple} = \mu_\text{PD}$, 
since the signal strengths are the same for these processes. 
Thus in Table (\ref{tab:lhc}), 
the processes in which we have to 
take into account the SM number of events $N^\text{SM}_\text{PD}$ 
are the $ZZ$ final states with ggF+ttH+bbH (ATLAS) 
and ggF+ttH (CMS). 
The integrated luminosities are 4.5 and 20.3 fb$^{-1}$ for  
$\sqrt{s}$=7 and 8 TeV respectively in the ggF+ttH+bbH (ATLAS), 
and 5.1 and 19.7 fb$^{-1}$ for ggF+ttH (CMS).

For the SM production cross sections, 
we use $\sigma_\text{SM}$ = (15.13, 1.222, 0.5785, 0.3351, 0.08632, 0.1558) pb 
\cite{SMHiggs_LHC7} for the (ggF, VBF, WH, ZH, ttH, bbH) processes respectively 
with LHC $\sqrt{s}$=7 TeV, 
and  $\sigma_\text{SM}$ = (19.27, 1.578, 0.7046, 0.4153, 0.1293, 0.2035) pb 
\cite{SMHiggs_LHC8} with LHC $\sqrt{s}$=8 TeV.
The SM cross sections for specific final states are calculated via 
$\sigma^\text{SM}_\text{PD}=\sigma^\text{SM}_{P}\text{BR}(h^0 \to D)$.

For these analyses combining different production channels, 
we assume the detector acceptances and efficiencies to be the same 
for different production channels and for different 
{center-of-mass} energies at {the} LHC. 
Thus,  when computing the signal strength $\mu_\text{multiple}$ for multiple processes, 
the detector acceptances and efficiencies 
($A\epsilon$) cancel in the numerator and the denominator of Eq.\ (\ref{eq:muth}) 
and do not affect the signal strength. 
To fully address the effects due to the detector acceptances and efficiencies
for different processes and colliding energies
requires a careful simulation, which is beyond the 
scope of this paper. 
We believe that neglecting the effects of the detector efficiency in the two analyses with 
$ZZ$ final states does not alter the conclusion of our analysis significantly.

To quantitatively evaluate how well a model fits the LHC data, we compute $\chi^2$ as follows 
\begin{equation}
\chi^2 = \sum_i \left( {{\mu_\text{th}^{i} - \mu_\text{data}^{i}} 
\over  \sigma_\text{data}^i    } \right)^2\,,
\end{equation}
where $\mu_\text{th}^i$ is the predicted signal strength in theory models, 
and $\mu_\text{data}^i$ and $\sigma_\text{data}^i$ are the signal strengths 
and the associated uncertainties for 
the decay and production channels given in Table (\ref{tab:lhc}).


\begin{table}[thbp]
\vspace{0.2cm}
\begin{center}
\begin{tabular}{c|c|c|c|c|c|c|c}
\hline
Model & Type & $\alpha$ & $\beta$ &  $\Delta M$ & $\chi^2_\text{ATLAS}$ & $\chi^2_\text{CMS}$ & $\chi^2_\text{LHC}$   \\\hline
&SM & - & - & - & 7.2 & 6.8 & 14.0   \\\hline
A &   I & -0.35 & 1.36 & 0 & {\bf 6.0} & 11.5 & 17.5   \\\hline
B &   I & -0.13 & 1.30 & 0 & {7.7} & {\bf 6.4} & 14.1   \\\hline
C &   I & -0.12 & 1.40 & 0 & {7.3} & 6.6 & {\bf 13.9}   \\\hline
E &   II & -1.45 & 1.44 & 0 & {\bf 4.3} & 15.7 & 20.0   \\\hline
F &   II & -0.21 & 1.38 & 0 & {10.0} & {\bf 6.3} & 16.3   \\\hline
G &   II & -0.166 & 1.40 & 0 & {6.5} & 7.3 & {\bf 13.8}   \\\hline\hline
M &   I & -0.26 & 1.23 & 5 & {6.8} & {6.8} & 13.7   \\\hline
N &   II & -1.11 & 0.49 & 2.7 & {6.6} & 6.5 & {\bf 13.0}   \\\hline
\end{tabular}
\caption{Benchmark models. $\Delta M$ in unit of MeV. 
}
\label{tab:benchmark}
\end{center}
\end{table}

The SM has $\chi^2=14$ with $\chi^2=7.2$ ($6.8$) for ATLAS (CMS) measurements. 
In the 2HDM-I with $\Delta M = 0$, the best-fit models give 
$\chi^2_\text{min}=6.0, 6.4, 13.9$ for ATLAS, CMS, and both measurements.  
In the 2HDM-II with $\Delta M = 0$, the best-fit models give 
$\chi^2_\text{min}=4.3, 6.3, 13.8$ for ATLAS, CMS, and both measurements.  
Thus, the 2HDM-I and 2HDM-II can provide a fit to data somewhat better than the SM. 
Several benchmark model points with vanishing and non-vanishing $\Delta M$  are provided in Table (\ref{tab:benchmark}).

We carry out a full $\chi^2$ analysis in the 2HDM parameter space spanned by $\alpha$ and $\beta$,  
\footnote{Plots with $\cos(\beta-\alpha)$ and $\beta$ (or $\tan\beta$) as axes are sometimes used. 
However, one model point in the parameter space spanned by $\cos(\beta-\alpha)$ and $\beta$ 
can correspond to two different model points in the 
parameter space spanned by $\alpha$ and $\beta$. 
Thus the interpretation on the variable $\cos(\beta-\alpha)$ is ambiguous and 
it should not be used.}
with both ATLAS and CMS data given in  Table (\ref{tab:lhc}).
The 1 $\sigma$ and 2 $\sigma$ contours are plotted with $\Delta\chi^2=2.3$ and 6.18, for 
the $\Delta M =0$ and  $\Delta M =3$ MeV cases, in Fig.\ (\ref{fig:chi2}). 
For the $\Delta M =0$  case, in both 2HDM-I and 2HDM-II models, the preferred model points reside in 
the large {$\tan\beta$} region. The sum of the decay widths {$\Gamma$} is less than 5 MeV
for the preferred 2HDM-I models, whereas in the 2HDM-II, the preferred models have 
15 MeV$\lesssim  \Gamma \lesssim$ 300 MeV. 
When $\Delta M =0$ is increased to $3$ MeV, the preferred 2HDM-I parameter space 
expands around its previously preferred regions; 
however, under the same $\Delta M$ increase in 2HDM-II, 
new parameter regions appear. 
We note in passing that the current Higgs mass measurement at the LHC 
has an uncertainty of $\sim$ 240 MeV \cite{Aad:2015zhl}, which, however, does not 
rule out the 2HDM-II models with $\Gamma>240$ MeV, 
because the analysis is carried out  
with SM assumptions and is not model-independent. 
Nevertheless, further reducing the mass uncertainties at the LHC can  
probe the preferred region of the 2HDM-II parameter space.

The interference effects between the two Higgs bosons depend strongly on the ratio $\Delta M/\Gamma$; 
when the mass separation $\Delta M$ becomes significantly large 
compared with the sum of the decay widths $\Gamma$, 
the interference effects begin to diminish, as illustrated in Fig.\ (\ref{fig:interference}). 
In the two types of 2HDM considered, the dependence of $\Gamma$ on the angle $\alpha$ is negligible; 
the dependence of $\Gamma$ on the 
$\beta$ angle is significant and can be well approximated as follows: 
$\Gamma/\Gamma_{h^0} \simeq 1 + (1-\text{BR}_{VV})/\tan^2\beta$ for the 2HDM-I, 
and $\Gamma/\Gamma_{h^0} \simeq 1 + (\text{BR}_{cc}+\text{BR}_{gg})/\tan^2\beta 
+ \tan^2\beta(\text{BR}_{bb}+\text{BR}_{\tau\tau})$ for the 2HDM-II, 
where the BRs denote the Higgs boson decay branching ratios in the SM.

To explore models with non-vanishing $\Delta M$, we carry out a three dimensional random scan 
with flat priors in 
{$ - \frac{\pi}{2} \leq \alpha \leq \frac{\pi}{2}$}, 
{$0 \leq \beta \leq \frac{\pi}{2} $}, and $\Delta M<100$ MeV. 
Fig.\ (\ref{fig:3d}) shows the 2HDM-II models that fall in the 1 $\sigma$ and 2 $\sigma$ region from the scan. 
These models all have $\tan\beta<10$, which corresponds roughly to
$\Gamma \lesssim 300$ MeV, consistent with the upper value 
in Fig.\ (\ref{fig:chi2}). 
There are two major bands of model points within  1 $\sigma$ and 2 $\sigma$ corridors in the random scan: 
$\beta=\alpha+{\pi\over 2}$, and $\beta=\alpha$. 
The former case has $\cos(\beta-\alpha)=0$ which is also known as the ``alignment'' limit, 
and the latter case has  $\sin(\beta-\alpha)=0$. 
In the former case, the coupling strength (normalized to SM values) 
between $h$ and SM particles are ($1,1,1,1)$ for 
(vector bosons, up-type quarks, down-type quarks, charged leptons), 
and the ones between $H$ and SM particles are ($0,1/\tan\beta, -\tan\beta, -\tan\beta)$ respectively. 
In the latter case, the coupling strength 
between $H$ and SM particles are ($1,1,1,1)$, 
and the ones between $h$ and SM particles are ($0,-1/\tan\beta, \tan\beta, \tan\beta)$, 
as in the order given before. These two bands are also exhibited in Fig.\ (\ref{fig:chi2}).

The best-fit model in the 2HDM-I prefers to go to much large $\Delta M$ values, such that 
our approximation no longer holds. Here
we provide a 2HDM-I benchmark model with small mass difference in Table (\ref{tab:benchmark}): 
model point M with $\Delta M = 5$ MeV. 
Unlike the 2HDM-I, the best-fit model in the 2HDM-II has $\Delta M \simeq 2.7$ MeV, which 
is represented by model point N in Table (\ref{tab:benchmark}).


\section{Higgs width and lepton collider}

The total decay width of the SM Higgs boson can be indirectly constrained by 
comparing the off-peak Higgs to $ZZ$ events with the on-peak ones \cite{Caola:2013yja}. 
However, this method can not be used to constrain the decay widths 
of the degenerate Higgs bosons, because this indirect method is based on the SM assumptions. 
A direct measurement of the Higgs width can be achieved in a muon 
collider by scanning the lineshape of the Higgs resonance, where 
one can unambiguously determine whether more than one 
scalar is present.    
It is anticipated that a 4 \% accuracy on the SM-like Higgs boson 
width can be achieved in a future muon collider with 0.003\% beam 
energy resolution and 1 fb$^{-1}$ integrated luminosity \cite{Han:2012rb}. 
At future $e^+e^-$ collider, the total Higgs width can be indirectly 
inferred via the $H\to WW$ process to a precision of (4-5)\% \cite{Brau:2012hv}. 
A clean and model-independent method to determine the Higgs mass is to 
reconstruct the recoil mass in the Higgs-strahlung process $e^+e^-\to h^0 Z$, 
which can probe the SM Higgs mass at the International Linear Collider (ILC), at future FCC-ee,  
and at the Circular Electron Positron Collider (CEPC). 
A precision of $\delta M_{h^0} \simeq 5.9$ MeV in the SM is expected {at the} CEPC \cite{cepc:preCDR}. 
However, the measurements at the {$e^+ e^-$} colliders discussed above are indirect and 
should be interpreted based on the underlying theory models; 
the precision discussed for the lepton collider are only applicable to a single SM Higgs boson case. 
Thus, the discrimination 
power regarding the degenerate Higgs bosons at electron collider is rather limited.

\section{Conclusion \label{sec:conc}}

In this work, we have analyzed the LHC signals arising from the scenario where two 
$CP$-even  Higgs bosons are almost degenerate in mass near 125 GeV. 
The quantum interference effects between degenerate Higgs bosons 
have to be taken into account in order to 
correctly predict the LHC signal, which, however, was often neglected in the literature. 
We have taken 2HDM-I and 2HDM-II as prototype models for our concrete numerical analysis; 
for certain regions in the parameter space, the interference effects can alter the predictions significantly.
We have carried out a global $\chi^2$ analysis using 27 measurements on Higgs signal strengths 
by ATLAS and CMS collaborations, which shows that the 2HDM with two degenerate 
$CP$-even Higgs bosons 
can fit the LHC data somewhat better than the SM. Future LHC data can improve the Higgs 
measurements with better accuracy, 
but it lacks the capability of distinguishing the two Higgs bosons that are degenerate in mass. 
It is also quite challenging to fully resolve this issue in the proposed {$e^+ e^- $} colliders. 
Nevertheless, for specific models, such as the 2HDM, analyses employing multiple collider searching channels 
with more LHC data (and/or data from future $e^+ e^-$ colliders) can further constrain the parameter space or even 
completely rule out the entire parameter space, but such results are model-dependent. 
To unambiguously discriminate the degenerate Higgs case from the single Higgs case, 
one may have to go to a future muon collider, where a scan around the Higgs resonance can 
be performed.

\section*{Acknowledgements}

We thank Yan-Wen Liu, Ning Zhou for helpful discussions. 
N.C.\ is partially supported by National Science Foundation of China (under Grant No.\ 11335007, 11575176), 
and the Fundamental Research Funds for the Central Universities (under Grant No.\ WK2030040069). 
The work of Z.L.\ is supported in part by the Nanjing University Grant 22721007 and  
the Tsinghua University Grant 523081007.

\section{ Appendix: The integral for $2\to 3$ phase space}

The signal strength in Eq.~\eqref{eq:muPD} is generalized from the results for the ggF production channel to two photons in Eq.~\eqref{eq:mu}.
Here, we prove that Eq.~\eqref{eq:muPD} is also valid for evaluating the signal strength for the $WH$ and $ZH$ processes.
Specifically, we consider the process of $q \bar q' \to V^* \to (h/H \to f \bar f) V$, whose parton-level amplitude reads
\begin{equation}
i{\mathcal M}\sim \bar v (p_1) \gamma^\mu ( g_V + g_A \gamma_5 ) u(p_2) \frac{-i }{\hat s - m_V^2 } \epsilon_\mu^*(k) 
\times \bar u (q_1) \sum_{A=h\,,H}\Big(  \frac{G_{AVV} }{q^2 - M_A^2 + i M_A \Gamma_A} G_{Aff}   \Big) v(q_2)\,.
\end{equation}
We find that
the integral of the amplitude squared over the $2\to 3$ phase space is reduced to the integral over the energy of the final-state vector boson $E_k$ as follows
\begin{equation}
\int | {\mathcal M} |^2 d{\rm PS}_{2\to 3} \sim \frac{1 }{ 4 m_V^2 (\hat s - m_V^2 )^2 } \int d E_k f(E_k) \sum_{A\,,B=h\,,H} \frac{G_{AVV} G_{Aff} }{E_k - E_k^{A*} } \frac{G_{BVV}^* G_{Bff}^* }{ E_k - E_k^B }\,,
\end{equation}
where 
\begin{subequations}
\begin{eqnarray}
f(E_k)&=& \sqrt{E_k^2  - m_V^2} \Big[  (\hat s - E_k \sqrt{\hat s} )( 2 E_k^2 + m_V^2 ) + 12 M_{\phi}^2 m_V^2 \Big] \,,\\
E_k^\phi&=& \frac{1}{2\sqrt{\hat s} } (\hat s + m_V^2 - M_\phi^2  -i M_\phi \Gamma_\phi ) \,.
\end{eqnarray}
\end{subequations}
Since the decay width is much smaller than the Higgs boson masses, $\Gamma_\phi \ll M_\phi$, only the $E_k$ values leading to the on-shell Higgs boson (i.e., $q^2= M_\phi^2$) contribute significantly to the integral. 
Thus, we can fix the $E_k$ values in the function $f(E_k)$ and move it outside of the integral such that
\begin{equation}
\int | {\mathcal M} |^2 d{\rm PS}_{2\to 3} \sim \frac{1 }{ 4 m_V^2 (\hat s - m_V^2 )^2 } f(E_k^0) \int d E_k \sum_{A\,,B = h\,,H} \frac{ G_{AVV} G_{Aff} }{ E_k - E_k^{A*}  } \frac{G_{BVV}^* G_{Bff}^* }{ E_k - E_k^{B} }
\end{equation}
The remaining integral can be computed via the residue theorem
\begin{equation}\label{eq:}
\int \frac{d E_k }{( E_k - E_k^{A*} ) ( E_k - E_k^B ) }  = \frac{2\pi i }{ E_k^{A*} - E_k^B } = \frac{4\pi i \sqrt{\hat s} }{ z_B^* - z_A } \,,
\end{equation}
which has the same dependence on $z_\phi$ as Eq.~\eqref{eq:ggtoaa_xsec}. Thus, we can obtain exactly the 
same expression for $\mu$ as in Eq.~\eqref{eq:mu} for the  $WH$ and $ZH$ processes.


\begin{thebibliography}{100}
\bibitem{custodial}   
The $W$ and $Z$ bosons share a common reduced coupling here due to
 the custodial symmetry in the 2HDM. 
If custodial symmetry 
gets broken in some other models, one has to distinguish the WBF and ZBF couplings 
in the VBF processes and also the WH and ZH couplings.
\end{thebibliography}
\end{document}